\documentclass[runningheads]{llncs}
\usepackage[T1]{fontenc}
\usepackage{graphicx}
\usepackage{placeins}
\usepackage[bookmarks=true,hyperfigures=true,colorlinks=true,linkcolor=blue,citecolor=blue,pdfpagemode=fullscreen,plainpages=false,pdfpagelabels]{hyperref}
\usepackage[all]{hypcap}
\usepackage{color}

\usepackage{url,cite}
\newcommand{\comment}[1]{}
\newcommand{\memo}[1]{}
\newcommand{\arxiv}[1]{\href{https://arxiv.org/abs/#1}{\tt arXiv:#1}}
\usepackage{wrapfig}
 \usepackage{fancyhdr}
\newcommand{\ifelse}[2]{#2}
 
\newcommand{\longver}[1]{\ifelse{}{#1}} 

\newcommand{\conf}{P_{\!\mathit{conf}}}
\newcommand{\vbar}{\,|\,}
\newcommand{\clasce}{{\sc Clarissa/asce}}
\begin{document}
\title{\ifelse{\vspace*{-2.7ex}}{}Confidence in Assurance 2.0 Cases}
\author{Robin Bloomfield\inst{1} \and 
John Rushby\inst{2}}
\authorrunning{R. Bloomfield, J. Rushby: \emph{Confidence in Assurance 2.0 Cases}}
\institute{Adelard (NCC Group), London N1 7UX England\\
and City, University of London\\
\email{r.e.bloomfield@city.ac.uk} \and
Computer Science Laboratory, SRI International\\ 
           Menlo Park, CA 94025 USA\\
\email{Rushby@csl.sri.com}}
\maketitle
\vspace*{-2.5ex}
\begin{abstract}
An assurance case should provide justifiable confidence in the
truth of a claim about some critical property of a system or
procedure, such as safety or security.  We consider how confidence can
be assessed in the rigorous approach we call Assurance 2.0.

Our goal is \emph{indefeasible} confidence and we approach it from four
different perspectives: logical soundness, probabilistic assessment,
dialectical examination, and residual risks.
\end{abstract}

\ifelse{}{\thispagestyle{fancy}}
\longver{\setcounter{footnote}{0}}

\vspace*{-5.5ex}
\section{Introduction}
\vspace*{-1.5ex}

In simple outline, Cliff Jones' work has focused on how we can build
trust in computer-based systems using mathematical reasoning.  As he
himself recognized, applying this as evidence to
justify dependability of critical system requires us to know how
much confidence we have in the reasoning, how it relates to properties
of the real world, and how it can be combined with other types of
evidence.  Assurance 2.0 provides a framework for answering these
questions.

\ifelse{ The popular view is that evidence is provided by testing so,
whenever there is a major systems failure, the first reaction of the
press and public is ``they didn't test it enough.''  But no feasible
amount of testing can provide the confidence required for safety and
other critical properties of computer-based systems.  If we test a
system for $n$ hours and observe no failures then, in the absence of
other information, the best prediction we can make is that the
likelihood of another $n$ hours with no failures is about 50\%
\cite[p.\ 73, and sidebar on p.\ 74]{Littlewood&Strigini93}.  Hence,
to secure assurance for failure rates of $10^{-9}$ per hour, as
required for many critical systems, we would need to test the system
for about $10^9$ hours, or around 115,000 years, which is completely
infeasible
\cite{Butler&Finelli93,Littlewood&Strigini93,Kalra&Paddock16:miles}.

But we do certify and deploy such systems and their safety record
justifies this, so how is it done?  The answer is that we do not test
``in the absence of other information'': thanks to analysis and reasoning, we have
prior confidence that the system harbors zero or very few critical
faults.  This confidence can be expressed as a subjective probability
so, if we are $95\%$ confident, we believe there is only 5\% chance
that the system contains critical faults.  We can now use
statistically valid random testing to explore the existence of
failures due to those potential faults but, unlike the previous case,
we know something about the system, so when we see $n$ hours with no
failures we can conclude (by what is called Conservative Bayesian
Inference, CBI \cite{Strigini&Povyakalo13,Zhao-etal:CBI20}) that we
are likely to see another $10 \times n$ with no critical failures.
Additional analysis based on operational experience 
\cite{Strigini-etal22:bootstrapping} further reduces
testing to feasible levels.
}
{Whenever there is a major systems failure, the first reaction of the
press and public is ``they didn't test it enough.''  But in fact, no
feasible amount of testing can provide the confidence required for
safety and other critical properties of computer-based systems.  For
example, in commercial airplanes, ``catastrophic failure conditions''
(those ``which would prevent continued safe flight and landing'') must
be ``so unlikely that they are not anticipated to occur during the
entire operational life of all airplanes of one type'' \cite{10-9}.
The ``entire operational life of all airplanes of one type'' is about
$10^{8}$ to $10^{9}$ flights for modern airplanes.  With an average
flight duration of about 90 minutes, this requires a critical failure
rate no worse than about $10^{-9}$ per hour.

If we test a system for $n$ hours and observe no failures then, in the
absence of other information, the best prediction we can make is that
the likelihood of no failures in another $n$ hours is about 50\%
\cite[p.\ 73, and sidebar on p.\ 74]{Littlewood&Strigini93}.  Hence,
to secure assurance for failure rates of $10^{-9}$ per hour we would
need to test the system for about $10^9$ hours, or around 115,000
years.  Even with 1,000 copies of the system on test, this is still
well over 100 years of continuous operation and is completely
infeasible
\cite{Butler&Finelli93,Littlewood&Strigini93,Kalra&Paddock16:miles}.

But we do assure and certify commercial aircraft and their safety
record justifies this, so how is it done?  The answer is that we do
not test ``in the absence of other information'': we have prior
confidence that the system harbors zero or very few critical faults.
This confidence can be expressed as a subjective probability so, if we
are $95\%$ confident, we believe there is only 5\% chance that the
system contains critical faults.  We can now use statistically valid
random testing to explore the existence of failures due to those
potential faults but, unlike the previous case, we know something
about the system, so when we see $n$ hours with no failures we can
conclude (by what is called Conservative Bayesian Inference, CBI) that
we are likely to see another $10 \times n$ with no critical failures
\cite{Strigini&Povyakalo13,Zhao-etal:CBI20}.  Another observation
\cite{Strigini-etal22:bootstrapping}, further reduces the testing
required so that $10^4$ hours might be adequate: on day 1, we do not
need assurance for the full lifetime of all airplanes of the type, we
will be satisfied with assurance for the first few months and the
first few airplanes, then we ``bootstrap'' our way forward by applying
CBI to the accumulating operational experience.
}

Central to this approach is prior confidence in the system (software)
concerned.  This confidence must be \emph{justified} so there has long
been interest in what evidence can provide adequate justification.
Correctness of software with respect to its specification has always
been seen as an element in this justification and by the 1970s there
were several methods and tools that attempted to mechanize reasoning
about software using formal specification languages and automated or
interactive theorem provers.  VDM was prominent among these and Cliff
Jones was a key member of the team at IBM Vienna Laboratory that
created it.  John Rushby, then at Newcastle and working on formal
assurance for computer security, recalls several visits to Oxford in
1979--80 where Cliff Jones was then based, with enlightening
discussions on formal methods and tools that contributed to his
decision to participate in development of {\sc Ehdm} and PVS, which
share similar goals to VDM\@.  During these discussions and activities,
it became apparent to us all that activities such as formal
verification and validation are important elements in achieving
justified confidence, but are not the whole story: we also need to be
sure that the properties verified are the right ones, that they are
stated correctly, that we can trust the methods used and any tools
employed, and so on.  As Cliff Jones stated \cite{Jones:tractable03}:
\begin{quote}
``\ldots one cannot repeat often enough that all that is even theoretically
possible is to prove\ldots that a program
satisfies a formal specification. Whether the formal text of the
specification actually does something useful is an issue which is not
susceptible to mathematical argument.'' 
\end{quote}

At approximately the same time, Robin Bloomfield was working for the
nationalized energy company CEGB on  how to assess
critical software proposed for nuclear reactor protection systems at
what became Sizewell B PWR\@.  With help from Cliff Jones, he and Peter
Froome developed VDM specifications for research reactor protection
systems, along with animations in Prolog.  This contributed to their work on
the overall problem of assurance and the developing concept of
a structured safety case.  Robin Bloomfield and Peter Froome founded
Adelard as a business to apply these ideas, with Cliff Jones on its
early advisory board, and together they presented industry courses on
these topics.

More recently, Robin Bloomfield and John Rushby have combined their
experience and worked to integrate formal methods with assurance
cases.  Together, they gratefully acknowledge Cliff Jones'
contribution to initiating and furthering their combined interest in
these topics, which were cemented by participating in the project
``Dependability of Computer-Based Systems'' that he led in  2001--7.

In this paper, we present the approach to justified confidence that we
call Assurance 2.0.  It employs several ideas that are not in
themselves new, but integrates them in a way that we believe is
coherent and effective.  Here, we mainly address traditional
computer-based systems, such as aircraft flight control, nuclear power
generation, etc.
We presume their developers consciously employ rational design
principles with an architecture and requirements matched to the environment, and
that design and assurance develop together.  The
new challenges of systems employing  AI or machine learning are
considered elsewhere
\cite{Bloomfield-etal:autonomy-templates21,Jha-etal:Safecomp20,Bloomfield&Rushby:AI24}.

The essence of ``justified confidence'' is that there must be
near-complete understanding of how the given system works and is
implemented, what are its hazards, how these are eliminated or
mitigated, and how we can be sure all this is done correctly.  The
purpose of an assurance case is to develop, document, challenge, and
communicate such justified confidence in critical properties of a
system or procedure, such as its safety or security.  We will speak
mainly of safety, but this should be understood to represent a
wide class of critical properties.

Assurance cases augment earlier
approaches to safety such as standards and guidelines by allowing more
choice in selection of techniques for ensuring and justifying safety.
Building on this, Assurance 2.0 is an approach to the development and
presentation of assurance cases that is intended to make their
construction and assessment more straightforward, yet also more
rigorous.  In fact, it is rigor that enables straightforwardness
because it reduces the ``bewilderment of choice'' and makes assurance
cases more systematic and predictable.  The present paper augments our
earlier ``manifesto'' \cite{Bloomfield&Rushby:Assurance2} by
describing how Assurance 2.0 cases can be evaluated and how confidence
in the case and in the system or procedure that it documents can be
assessed, and it also indicates how automated assistance for these
activities is provided and applied in the \clasce\ toolset developed
with colleagues at Honeywell and UT Dallas
\cite{Varadarajan-etal:DASC24}.  A small but realistic example
application of Assurance 2.0 and its \clasce\ tool is provided
elsewhere \cite{Bloomfield-etal:Hardens24}, and others are in
development. 

It should be noted that those who assess and certify a system have a
different view on its assurance case than those who develop it.  The
task of the developers is to construct a system that is safe together
with an assurance case that provides true and compelling reasons for
believing it is so.  The task of the assessors is to make sure that
the developers have accomplished this and to be confident in accepting
or rejecting the system: they do not repeat the work of the developers
and reconstruct the assurance case, they review it.  However, they may
commission independent and diverse analyses that could generate new
evidence and possibly revise the case, and they may prepare their own
``sentencing statement'' \cite[Section
7]{Bloomfield&Rushby:confidence22}.  \longver{ The exact form of the
assessment and review varies according to the system and its critical
properties, and the industry concerned: assessment for nuclear power
generation is different than for civilian aircraft.  } Here, we focus
on the developers' viewpoint, although we do indicate how some
capabilities should be useful to assessors.  A paper on the assessor's
viewpoint and the notion of a ``metacase'' (a case about the case) is
in preparation. 

Confidence in Assurance 2.0 has four components: two positive views
(logical and probabilistic) that assess the extent to which an
assurance case sustains the claimed properties, and two negative views
(defeaters and residual risks) that explore doubts and challenges and
the potential impact of any doubts that remain.  The four components
are described below, each in a separate section, followed by a summary
and conclusion.

\section{Logical Assessment}
\label{logic}

An assurance case in Assurance 2.0 is composed of \emph{claims},
\emph{argument}, and \emph{evidence}.  Claims are statements, usually
in natural language, about some property of the system or its design,
development, construction, environment, etc.  Many of these are
described by \emph{models} and much of assurance is about
relationships among them (e.g., do the specifications accurately
reflect the requirements?) and about establishing a path from high-level to
lower-level models and ultimately to the actual system.  Evidence
refers to observations, measurements, or experiments on the system or
its (models of) design, development, construction, environment and so
on, that justify certain claims.  The overall argument is a collection
of \emph{reasoning} (or argument)\footnote{``Argument'' is an
overloaded word: we generally use it to refer to the totality of
claims, reasoning steps, and evidence.  Likewise, ``case'' refers to
the argument plus the totality of all other material submitted and
developed for and during assessment.}  \emph{steps} that each justify
a parent claim on the basis of some child claims (we usually call them
\emph{subclaims}) or that introduce evidence.
In total, the reasoning and evidential
steps build a \emph{structured argument} from the evidence to a
\emph{top claim} that states the critical property to be assured.  The
argument can be displayed graphically as a tree-like
structure\longver{ such as shown in Figure \ref{traffic},} where differently
shaped nodes are used to indicate claims, evidence, and reasoning
steps.  The argument structure must be a connected graph but might not be a
true tree because some claims may serve as subclaims to more than one
reasoning step.

A critical element associated with each reasoning and evidential step is a narrative
\emph{justification} that provides a compelling explanation why the
parent claim follows from its subclaims or its evidence.  The justification may
reference a \emph{side-claim} that logically functions like a subclaim but has a
conceptually different r\^{o}le that is discussed later.  The leaves
of the argument structure must be either evidence, \emph{assumptions}
(which are claims that are
specially designated as such), references to external
\emph{subcases} developed (or to be developed) separately, or
\emph{defeaters} that have been accepted as \emph{residual doubts} (see Sections
\ref{defeaters} and \ref{residuals}).

An argument in Assurance 2.0 is interpreted as a logical proof from
evidence and assumptions to the top claim, where, in each reasoning
step, the conjunction of the side-claim and subclaims deductively
entail the parent claim.  The reasoning steps function as \emph{a
priori} premises, meaning that we believe them by virtue of
understanding the justifications supplied for each of them.  This is
in contrast to evidence incorporation steps, which are \emph{a
posteriori} premises, meaning that our belief in their claims rests
on the evidence supplied.

This style of argument is called \emph{Natural Language Deductivism}
(NLD). It is an informal counterpart to deductive proof in formal
mathematics and logic but differs in that its premises are
``reasonable or plausible'' rather than certain, and hence its
conclusion (i.e., top claim) is likewise reasonable or plausible
rather than certain.  NLD differs significantly from other
interpretations of informal argumentation (as used in debate, for
example), where weaker or different forms of inference may be used
\cite{Blair15}; indeed, the very term ``natural language deductivism''
was introduced as a pejorative to stress that this style of argument
does not truly represent ``informal argument'' \cite{Govier87}.
However, our focus is not informal arguments in general, but the
structured arguments of assurance cases, and so in Assurance 2.0 we
depart from the previous association of assurance cases with
``informal argument'' and adopt the label NLD with pride.

We also depart from other treatments of assurance cases by raising the
bar on interpretation of ``reasonable or plausible'' and require the
argument to be \emph{indefeasible}.  This is a criterion from
epistemology and means that the overall argument, and its evidence and
justifications in particular, must be so compelling, and all credible
doubts and objections must have been so thoroughly considered and
countered, that we are confident none remain that could change the
decision \cite{Rushby:Shonan16}.  This does not mean that we must
eliminate all doubts, but that we have identified them and are
confident they will not change the decision (see Section
\ref{residuals} on residual doubts).  In addition to defining the
assessment criterion, indefeasibility provides the ``stopping
condition'' for both development and assessment of an Assurance 2.0 case.

The world is uncertain and our understanding imperfect, so
indefeasibility is very demanding.  That is why we consider it
necessary to examine and evaluate an assurance case from several
diverse perspectives whose combination can yield an overall assessment
of its indefeasibility.  Some perspectives focus on ``positive''
aspects of the case, such as the evidence and argument in support of
its claims, while others consider the ``negative'' aspects (i.e., its
potential defeasibility), such as doubts and objections (represented
as defeaters) that have been considered and refuted, and any 
that remain as residual doubts.

The reason we opt for NLD and indefeasibility is that anything less
must have gaps or errors that may mask a safety flaw.  This is
particularly so for nondeductive (sometimes called ``inductive'')
reasoning steps: if the subclaims to a step do not deductively entail
the parent claim but merely ``strongly suggest'' it, then either the
step is fallacious or there are missing subclaims.

An important purpose of side-claims in reasoning steps is to factor
justification for deductiveness.  For example, a step may divide the
parent claim into subclaims over some explicit enumeration, such as
components of the system, or time (e.g., past, present, future), or
hazards, and so on.  In each case, the side-claim must establish that
the decomposition satisfies any properties that may be needed to make
it deductive, such as the subclaims partition the parent claim, or the
claim distributes over components, or that some theory justifying the
decomposition is properly applied.  By factoring these concerns into
the side-claim, we give them the focus they require while also
simplifying the core argument.  A side-claim may be justified by a
subargument of its own, or it may become the top claim of a separate
subcase, or it may be left unsupported as an explicit assumption, and
it can be challenged by a defeater (see Section \ref{defeaters}).

\longver{For example, if the
decomposition is over hazards, then the side-claim will require that
all hazards have been identified and that the decomposition considers
them all, both individually and in combination; such a side-claim
might be discharged by evidence that attests to use of a well-accepted
method of hazard analysis, performed diligently.  In other cases, the
side-claim should ensure that the subclaims partition the parent claim
(i.e., no overlaps).  Each of the other four kinds of argument blocks
(see later) may have a side-claim, specific to its kind and its particular
application.}

NLD is practical in Assurance 2.0 because its argument does not
perform complex logical reasoning such as that required to verify an
algorithm or to prove that a specification satisfies its requirements:
these are regarded as calculations that should be performed
externally, with their results provided to the assurance case as
evidence.  Thus a claim is typically an atomic proposition such as
``source code \texttt{x} satisfies specification \texttt{y}'' that
is justified by evidence ``formally verified by \texttt{z}''
together with narrative documentation of the verification performed.

When augmented with supporting claims about methods and tools used,
why they can be trusted, their provenance etc., the ``satisfies''
claim and its evidence become a self-contained subcase.  In generic
and possibly parameterized form, subcases that deal with standard
assurance topics such as satisfaction/refinement between requirements and
specifications, use of static analysis to guarantee absence of runtime
exceptions, application of MC/DC testing and so on, become what we call
\emph{assurance theories} and an Assurance 2.0 argument mainly
assembles applications of such theories.  Specifically, large parts of
the argument will be synthesized by instantiating pre-existing (and
ideally ``pre-certified'') theories, other parts may be new ``bespoke''
construction (although we recommend these are first developed as
reusable theories), and some will be ``glue logic'' that ties all the
parts together into a coherent whole.

In addition to reducing duplication of effort and eliminating
``homespun'' treatments, pre-existing theories assist in communication
and comprehension of assurance arguments: the high-level structure of
an argument can be indicated by enumerating the theories that it uses.
Dually, much of an assurance case can be synthesized automatically
using a library of theories as ``templates.''  \clasce\ has a
``synthesis assistant'' that does this \cite{Bloomfield:CLARISSA-SA}.

\begin{figure}[h]
\vspace*{-4ex}
\center
\includegraphics[width=1.0\linewidth,trim=3cm 0cm 0cm 3cm, clip=true]{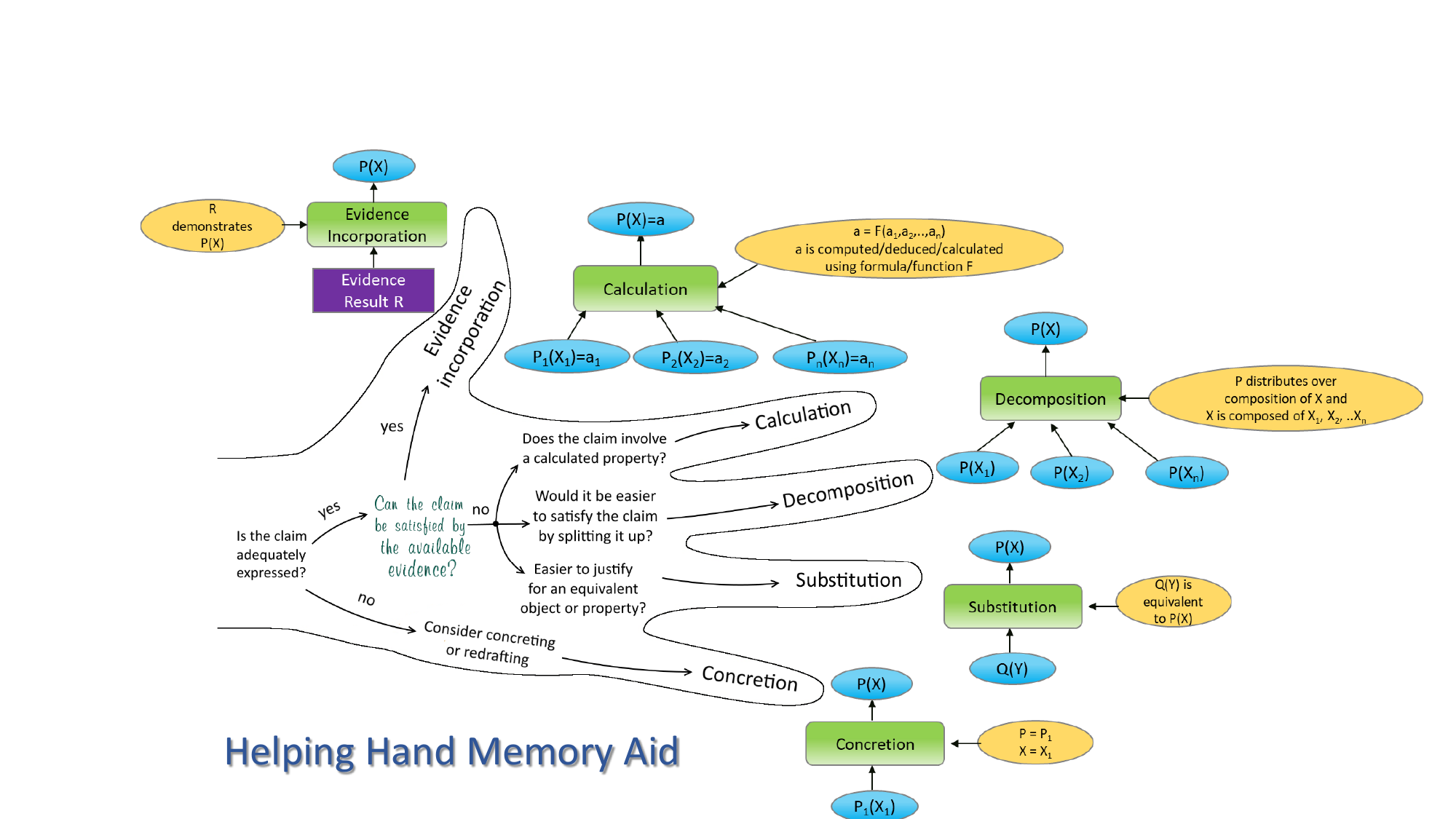}
\vspace*{-4ex}
\caption{Assurance 2.0 Building Blocks and ``Helping Hand'' (from \cite{Varadarajan-etal:DASC24})}
  \label{blocks-hand}
\vspace*{-4ex}
\end{figure}

\longver{
\begin{figure}[t]
\begin{center}
\includegraphics[width=1.0\textwidth]{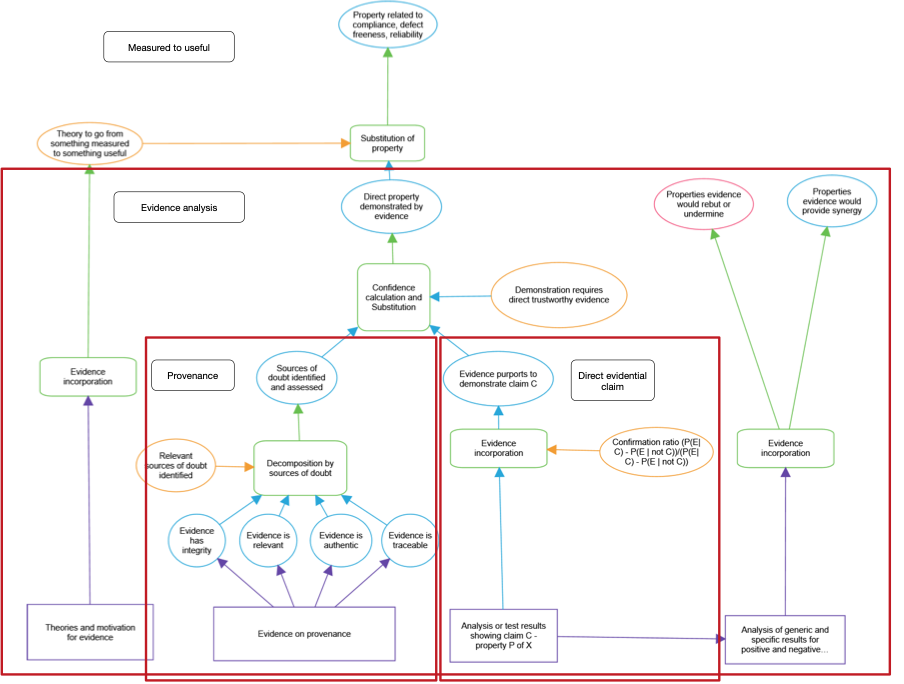}
\end{center}
\vspace*{-2.5ex}
\caption{\label{traffic}Portion of \clasce\ Canvas}
\vspace*{-2.5ex}
\end{figure}
}

Just as theories systematize the macro-scale structure of Assurance
2.0 arguments, so we also systematize the micro-scale.  Traditional
assurance case arguments are ``free-form'' in that reasoning steps can
have any ``shape'' and justification.  In Assurance 2.0, we restrict
reasoning steps to five ``building blocks'' (which we generally
abbreviate to \emph{blocks}) \cite{Bloomfield&Netkachova14}.  These
are \emph{decomposition}, \emph{substitution}, \emph{concretion},
\emph{calculation}, and \emph{evidence incorporation}.
Figure \ref{blocks-hand} shows the ``shape'' of each of the
five blocks together with the ``helping hand'' mnemonic that provides
guidance on their \ifelse{use.}{use,
and Figure \ref{traffic} illustrates the structure of a simple case.}
In the typical
structure of an Assurance 2.0 argument, general claims at the top are
refined into more precise claims using concretion steps, then
substitution steps are used to elaborate these claims about high level
models into claims about lower level models and their implementations.
The lowest-level claims in this structure must be assumptions (or
residual doubts)
or be
discharged by evidence.  Application of evidence is generally
accomplished in two steps: the lowest step performs evidence
incorporation to justify a claim about ``something measured'' (e.g.,
``we did requirements-based testing and achieved MC/DC coverage''
or ``we performed a requirements review'') and
this supports a second step, which is usually a substitution based on
application of an external theory, that connects it to a claim about
``something useful'' (e.g., ``we have no unreachable code''
or ``the
requirements correctly describe the desired
behavior'').  At any
stage, the argument may divide into subcases using decomposition or
calculation steps that enumerate a claim over some assembly (e.g.,
over components, requirements, hazards, etc.) or that split the
conjuncts of a compound claim.  This argument structure may recurse within
subcases.  

Now that we have some understanding of the purpose, structure, and
assessment goals for an Assurance 2.0 argument, we describe how logic
and epistemology are used to perform a positive assessment.  We said
earlier that an NLD argument is basically a proof of the top claim
using the reasoning steps and grounded on evidence.  Hence, the
criteria we apply are those from logic: validity and soundness.
Validity says the argument ``makes logical sense,'' assuming its
premises (i.e., reasoning and evidential steps) are true, no matter
what the claims mean (strictly, it is true in all interpretations).
Soundness says that in addition the premises are true for the actual
claims.  Because the structure of an Assurance 2.0 argument is so
restricted (its logical interpretation is simply propositional
calculus over definite clauses), validity reduces to a structural
check: viewed as a graph, the argument must be connected, with a
single top claim, and its leaves must be evidence incorporation nodes
or assumptions.  Subcases may either be expanded in place and
assessed as part of the main argument (i.e., as ``macros'') or be
separately assessed and represented by their own top claim (i.e.,
they function as lemmas).  Notice that this restricted argument
structure, which is enforced by  \clasce,  eliminates many common
errors and ensures that the argument is logically valid ``by
construction.''  Validity checking becomes more complicated in the
presence of defeaters and we defer detailed description to
Section \ref{defeaters}.

Given a valid argument, soundness adds the requirement that its actual
premises must be true.  In an Assurance 2.0 argument, the premises are the
evidence incorporation steps and the reasoning steps.  As fallible
humans, we cannot know that these are true, but we can attempt to
establish indefeasible belief that they are so.  We now describe how
this is done in Assurance 2.0.

\subsection{Confidence in Evidence}
\label{confirmation}

\ifelse{ As we mentioned earlier, it takes two steps to deploy
evidence in an Assurance 2.0 argument.  The first step uses an
evidence incorporation block to establish that the evidence supports a
claim about ``something measured'' that describes what the evidence
\emph{is}.  The second step substitutes the measurement claim by one
about ``something useful'' that describes what the evidence
\emph{means}.  The transition between the two typically uses a
substitution block that cites some external theory (which may have its
own assurance case) to justify the substitution.  }{ As we mentioned
earlier, it takes two steps to deploy evidence in an Assurance 2.0
argument.  The first step uses an evidence incorporation block to
establish that the evidence supports a claim about ``something
measured'' that describes what the evidence \emph{is}.  For example,
it could be a claim that a certain type of testing was performed and
what its results were (notice that it is test \emph{campaigns} that
constitute evidence, not individual tests), or it could be a similar
type of claim about static analysis, or a requirements review.  The
second step substitutes the measurement claim by one about ``something
useful'' that describes what the evidence \emph{means}.  For example,
it could be a claim that there is no unreachable code, or that the
code will have no runtime exceptions, or that the requirements
correctly describe the desired behavior.  The transition between the
two claims typically uses a substitution block that cites some
external theory (which may have its own assurance case) to justify the
substitution.  }

Basic assessment of the two steps requires examination of the
narrative justifications provided for the evidence incorporation and
substitution blocks, together with the evidence, claims, and external
theories used.  Notice that evidence must usually be packaged as an
\emph{evidence assembly} that includes descriptions of its provenance,
results obtained, methods and tools used, etc.

Basic assessment, however, merely provides confidence that the cited
evidence does indeed support the measured and useful claims: it does
not indicate how persuasive it is nor how much \emph{confidence} we
should have in its support for the useful claim.  Evidence is a bridge
from the world of observation and measurement to the world of claims
and logic.  Therefore we cannot assess persuasiveness of evidence by
the methods of logic, we need the methods of epistemology.
Epistemology is about justified belief (as an approximation to truth)
and it is natural to express the strength of our confidence in a
belief as a number.  We will expect those numbers to obey certain
rules (the Kolmogorov Axioms) and so they function as subjective
probabilities.

A natural measure of confidence in a claim $C$ given the evidence $E$
is the subjective posterior probability $P(C \vbar E),$ which may be
assessed numerically or qualitatively (e.g., ``low,'' ``medium,'' or
``high'').  However, confidence in the claim is not the same as
confidence that it is justified by the evidence.  It is possible that
the reason for a high valuation of $P(C \vbar E)$ is that our prior
estimate $P(C)$ was already high, and the evidence $E$ did not
contribute much.  So it seems that to measure justification we ought
to consider the difference from the prior $P(C)$ to the posterior $P(C
\vbar E)$ as an indication of the ``weight'' of the evidence $E$\@.
Difference can be measured as a ratio, or as arithmetic difference.
For brevity, we will use only ratio measures here; difference and
other measures are described in our report
\cite{Bloomfield&Rushby:confidence22}. 
An example of a ratio \emph{confirmation measure} is due
to Keynes:
\vspace*{-1ex}
\begin{eqnarray*}
\mbox{Keynes}(C, E) & = & \log\frac{P(C \vbar E)}{P(C)}.
\end{eqnarray*}
The logarithm (which may use any positive base) in Keynes' and other
ratio measures serves a normalizing purpose so that, as with
arithmetic difference, positive and negative confirmations correspond
to numerically positive and negative measures, respectively, and
irrelevance corresponds to a measure of zero.

There are many other confirmation measures proposed in the literature
\cite{Tentori-etal07}.  For example, some prefer to use the likelihood
$P(E \vbar C)$ rather than the posterior $P(C \vbar E),$ because it is
generally easier to estimate the probability of concrete observations
(i.e., evidence), given a claim about the world, than vice-versa,
thereby giving us the likelihood variant of Keynes'
measure:\footnote{Observe that Bayes' Theorem gives $\mbox{L-Keynes}(C, E) =
\mbox{Keynes}(C, E)$; other measures also have arithmetic relationships
between their posterior and likelihood variants.}
\begin{eqnarray*}
\mbox{L-Keynes}(C, E) & = & \log\frac{P(E \vbar C)}{P(E)}.
\end{eqnarray*}
We can see that these measures will tend toward their maxima when
$P(E)$ is small, meaning that $E$ should be unlikely in general.  This
suggests that we should favor evidence whose occurrence (in the
absence of $C$) would be a \emph{surprise}.
Similarly, if we have accepted evidence $E_1$ and seek additional
evidence, we should look for $E_2$ that is (or remains) surprising in
the presence of $E_1.$  Thus, for example, if $E_1$ is evidence of
successful tests, it will not be surprising if additional tests are
successful; instead we should seek evidence $E_2$ that is ``diverse''
from $E_!,$ such as static analysis\ifelse{ (see our report
\cite{Bloomfield&Rushby:confidence22} for more detail).}{.
More formally,\footnote{We use $\supset$ for material implication, $\wedge$
for conjunction, $\neg$ for negation, $\equiv$ for equivalence, and
$\approx$ for approximate (numerical) equality.} we have, by the
chain rule
\begin{eqnarray*}
P(C \wedge (E_2 \wedge E_1)) & = & P(C \vbar E_2 \wedge E_1) \times
    P(E_2 \vbar E_1) \times P(E_1), \mbox{\ and}\\
P(E_2 \wedge (C \wedge E_1)) & = & P(E_2 \vbar C \wedge E_1) \times P(C \vbar E_1) \times P(E_1).
\end{eqnarray*}
The left (and hence right) hand sides are equal, and so
\begin{equation}
\frac{P(C \vbar E_2 \wedge E_1)}{P(C \vbar E_1)} =  \frac{P(E_2 \vbar
    C \wedge E_1)}{P(E_2 \vbar E_1)}.\label{diverse}
\end{equation}
Thus, $E_2$ delivers the largest ``boost'' to Keynes' measure for the
justification provided by $E_1$ (i.e., the left hand side of
(\ref{diverse})) when $E_2$ would be surprising given only $E_1,$ but
not when given $C$ as well, which confirms that $E_2$ should be
\emph{diverse} from $E_1.$
These observations about ``surprising'' and ``diverse'' evidence are
intuitive, but it is satisfying to see them put on a
rigorous footing.
}

An additional consideration when evaluating evidence is that
it is not enough for the evidence to
support a given claim $C,$ it should also discriminate between that
claim and others, and the negation, or ``counterclaim'' $\neg\, C$ in
particular.  Again, discrimination or distance can be measured as a
ratio or as arithmetic difference; here, we use the ratio measure due to
Good (and Turing) from codebreaking activities during World War 2:
\vspace*{-2ex}
\[\mbox{Good}(C, E) = \log\frac{P(E \vbar C)}{P(E \vbar \neg\, C)}.\]

\longver{
We will refer to these as ``Type 2'' confirmation measures, and the
previous examples as ``Type 1.''  However, likelihoods are related to
posteriors by Bayes' Rule, and appearances of $P(\neg\, x)$ in Type 2
measures can be replaced by $1 - P(x)$ and then the distinction
between Type 2 and Type 1 measures disappears.   Manipulations of this
kind yield
\vspace*{-2ex}\begin{eqnarray*}
\mbox{Good}(C, E) & = & \log\frac{O(C \vbar E)}{O(C)}
\end{eqnarray*}
where $O$ denotes \emph{odds} (i.e., $O(x) = P(x)/(1-P(x))$) and
Good's measure is therefore sometimes referred to as the ``log odds''
or ``log odds-ratio'' measure for weight of evidence
\cite{Good:weight83}.
Similar manipulations of other confirmation measures 
reveal that they generally satisfy the following conditions:
\begin{enumerate}
\item They can be expressed as functions of $P(C \vbar E)$ and $P(C)$ only,

\item They are increasing functions of $P(C \vbar E),$ and

\item They are decreasing functions of $P(C).$

\end{enumerate}

Not all confirmation measures satisfy 1 above.  For example, a measure
due to Carnap, $P(C \wedge E) - P(C)\times P(E),$ depends nontrivially
on $P(E).$ However, such measures can be manipulated by irrelevant
evidence \cite[section 2]{Atkinson12}, so we prefer measures that
satisfy condition 1.
}

Although all confirmation measures assess the degree to which
evidence $E$ justifies claim $C,$ they do so in different ways and we
may prefer one measure to another (or prefer different measures for
different purposes) \cite{Hajek-Joyce:confirmation08}.  In particular,
different measures are based on elicitation of different judgments and
we believe there can be value in asking those who assess evidence to
consider the different points of view underlying these different
measures.  For example, $\mbox{Keynes}(C, E)$ elicits judgments $P(C
\vbar E)$ and $P(C)$, while $\mbox{L-Keynes}(C, E)$ elicits $P(E \vbar
C)$ and $P(E)$.  Furthermore, the two measures should yield the same
numerical value; we can therefore provide feedback to assessors if their
judgments are inconsistent.  Similarly, \longver{the original
formulation of} $\mbox{Good}(C, E)$ elicits judgment of $P(E \vbar
\neg\,C)$, which requires consideration of a contrary point of view.

\longver{
Note that different confirmation measures are not ordinally
equivalent.  That is to say, a given measure may rank one scenario
(i.e., combination of $P(C \vbar E)$ and $P(C)$) higher than another,
but a different measure may do the reverse.  However, in assurance we
are interested in strong confirmation and all measures are likely to
agree in these cases.  (A fourth condition can be added to the list
above to yield what are called \emph{justification measures} and all
of these are ordinally equivalent \cite{Atkinson12,Shogenji12}.)
}

Some will be skeptical that human developers and evaluators are able
to assess and manipulate probabilistic measures correctly, even
qualitatively, and will also contend that confirmation measures are
beyond everyday experience.  They may point to alleged flaws in human
evaluation of probabilities.  Here is a standard illustration
\cite{Tversky&Kahneman:conjunction83}.
\begin{description}
\item[Evidence $E$\,:]
Linda is 31 years old, single, outspoken and very
bright. She majored in philosophy.  As a student, she was deeply
concerned with issues of discrimination and social justice, and also
participated in anti-nuclear demonstrations.
\end{description}
\noindent
The challenge is to assess which of the following two claims is most
probable, which we interpret as best
supported by the evidence.
\begin{description}
\item[Claim $C_1$:] Linda is a bank teller,

\item[Claim $C_2$:] Linda is a bank teller and active in the feminist movement.
\end{description}

When human subjects are exposed to this and similar examples, they
overwhelmingly favor $C_2.$ Psychologists label this the ``conjunction
fallacy'' because $C_2$ is the conjunction of $C_1$ with another
clause and a conjunction must always be \emph{less} probable than
either of its components; they then cite this as evidence 
that intuitive human reasoning is poor at probabilities
\cite{Tversky&Kahneman:conjunction83,Kahneman11}.  However, a more
recent interpretation is that humans evolved to weigh evidence and
actually base their judgments on subconscious mental measures more
akin to confirmation than simple probabilities (even when asked about probabilities)
\cite{Crupi-etal08,Shogenji12}.To see this, we observe that the evidence $E$ seems to add nothing to
our prior belief in $C_1$ but it does seem to support the second
clause of claim $C_2$ (i.e., ``is active in the feminist movement'')
and so\longver{, by item 2 of the list of properties for confirmation measures,}
we can conclude that the evidence indeed confirms $C_2$ more than $C_1,$
thereby refuting the ``fallacy'' charge.

Our technical report \cite{Bloomfield&Rushby:confidence22} provides
more examples and several other (we think, interesting) illustrations
why confirmation
measures are appropriate and useful tools in assessment of evidence,
and also correspond to natural human judgment.  We also provide
further examples on the significance of the distinction between
measured and useful evidential claims and on selection of effective
evidence.

\ifelse{ In Assurance 2.0 we advocate that assessment of evidence
should make use of confirmation measures applied to the ``something
useful'' claim of the evidence concerned.  We want evidence that has
large positive confirmation measures and we recommend use of diverse
measures as this can probe the r\^{o}le of the evidence in both
supporting and undermining its evidential claims.  The measures can be
applied with several levels of ``rigor.''  At one extreme we can
estimate values for the constituent probabilities employed (e.g., by
asking experts to set a pointer on a scale) and evaluate the measures
numerically, while at another we can just use the ``ideas'' and think
informally about the relationships involved.  At an intermediate level
we can perform ``qualitative'' estimation and evaluation (e.g.,
``low,'' ``medium,'' or ``high'').  The \clasce\ tool has widgets
that can assist the
calculation and assessment of confirmation measures.  }{}

\longver{A different example that highlights the importance of
selecting evidence appropriate to the claims it is intended to support
is the ``Paradox of the Ravens'' \cite{Hempel45}.  Here, we seek
evidence for the claim ``all ravens are black''; the equally valid
contrapositive of this claim is ``all non-black objects are
non-ravens'' for which a white shoe is produced in evidence, allowing
the triumphant declaration ``that proves it: all ravens \emph{are}
black!''

A rational escape from this ``paradox'' is \emph{Nicod's criterion}
\cite{Nicod30} that only observations of ravens should affect our
judgment whether all ravens are black.  More generally, claims about
some class of objects can be confirmed or refuted only by evidence
about those objects.  Under this criterion, we expect that
observations of black ravens would tend to confirm our claim, while a
non-black raven definitely refutes it.  Good, in a cleverly titled
one-page paper \cite{Good:red-herring67}, rebuts this expectation with
an example where observation of a black raven disconfirms the claim
``all ravens are black.''

Examples such as this are challenging to philosophers seeking to
explain and justify the methods of science, but for assurance the
salient points are that we need to be skeptical about evidence (hence
consideration of alternative claims and counterclaims) and may need to
collect additional evidence to rule out alternative explanations.  (In
Good's example, observations of additional birds would allow us to
discriminate between the hypothesized situations.)  Confirmation
measures provide an attractive framework in which to probe these
issues and, far from being difficult for human evaluators, they
correspond to inbuilt human faculties for the weighing of evidence.

However, there are some further complications.
In the Linda example, claim $C_2$ entails the further claim ``Linda
works outside the home'' (since she is a bank teller), but the
evidence provides no direct support for this and it could easily be
false.  Thus, we have evidence that soundly supports a claim that
logically entails a further claim, yet that second claim could be
false.  For a more extreme example, the evidence that a card drawn
from a deck is an Ace supports the claim that the card is the Ace of
Hearts, and this entails the further claim that the card is red.  But
the card used in evidence could have been the Ace of Clubs, which
refutes the derived claim that it is red.

A pragmatic resolution to this apparent paradox is that if an
evidentially supported claim is a conjunction (e.g., card is both Ace
and Hearts), then we need indefeasible support for all elements of the
conjunction, and so, in the context of assurance, we should not accept
that the evidence about Linda is sufficient to establish claim $C_2$
(because it does not establish its $C_1$ conjunct).  These examples
also illustrate significance of the measured vs.\ useful distinction
on evidential claims, with confirmation measures applied to the
latter.  ``Card is an ace'' is a measurement claim, and is justified,
by the evidence, whereas ``card is red'' is a useful claim and any
confirmation measure will indicate that the proffered evidence is
irrelevant to this claim.

Sometimes a mismatch between evidentially measured and useful claims
leads to the realization that one or the other is misstated.  For
example, during World War 2, the US Army Air Force came to its
Statistical Research Group in New York seeking advice on where best to
add armor to improve the survival of their airplanes.  Many damaged
planes returning from engagements had been examined and this produced
the evidence shown below.

\begin{wrapfigure}{R}{.5\textwidth}
\vspace*{-6ex}
\begin{center}
\begin{tabular}{|l|l|}
\hline
Section of plane & Bullet holes per sq. ft. \\
\hline
Engine & 1.11 \\
Fuselage & 1.73 \\
Fuel system & 1.55 \\
Rest of plane & 1.8 \\
\hline
\end{tabular}
\end{center}
\vspace*{-6ex}
\end{wrapfigure}

The fuselage seems the most heavily damaged of the identified
components, so the evidence seems to support the claim ``the place
where armor will best improve survival of the plane is the fuselage.''
This is actually an evidentially useful claim; we should begin by
using the evidence to justify a measurement claim.  A plausible
candidate for this is ``the fuselage is the part of the plane with
heaviest damage.''  An important difference in these two claims is
that the measurement claim speaks of ``damage'' while the useful claim
speaks of ``survival.''  Thus we need either some inference from
damage to survival, or the measurement claim should also speak of
survival.  This leads to a key insight: the evidence comes exclusively
from planes that survived.  Hence the measurement claim should be
changed to ``the fuselage is the part of the plane that can survive
heaviest damage.''  From there it is a short step to deduce that
planes with heavy damage to the engines did not survive and hence the
celebrated advice of Abraham Wald that the best place to apply
armor is where there are no bullet holes \cite{Ellenberg15}.

We have now seen several examples why Assurance 2.0 attaches
importance to the distinction between measured and useful evidential
claims and to the use of confirmation measures in assessing the
relevance of the measured evidence to the evidentially useful claim.
We want evidence that has large positive confirmation measures and we
recommend use of diverse measures as this can probe the r\^{o}le of
the evidence in both supporting and undermining its evidential claims.
Confirmation measures can be applied with several levels of ``rigor.''
At one extreme we can estimate values for the constituent
probabilities employed (e.g., by asking experts to set a pointer on a
red/white/green scale) and evaluate the measures numerically, while at
another we can just use the ``ideas'' and think informally about the
relationships involved.  At an intermediate level we can perform
``qualitative'' estimation and evaluation.  For example, we could
elicit expert opinion on constituent probabilities according to a
five-level qualitative scale: \emph{certain} (this is true),
\emph{very confident}, \emph{confident}, \emph{neutral},
\emph{surprised} (if this is true), and \emph{very surprised}.  The
confirmation measures can then be evaluated using representative
numerical assignments to these estimates: for example, 1.0, 0.99, 0.9,
0.5, 0.1, and 0.01, respectively.  Thus, if an expert is initially
\emph{neutral} on a claim, but becomes \emph{confident} when given the evidence,
Keynes measure evaluates to 0.26.  This is rather lukewarm, so we
could then elicit the L-Keynes or Good measures in order to probe the
discriminating quality of the evidence.  The \clasce\ tool has some
widgets (illustrated in Figure \ref{widgets}) that can assist the calculation and assessment of confirmation
measures in this way.
}

\subsection{Confidence in Reasoning Steps}
\label{validity}

We noted earlier that an assurance argument in Assurance 2.0 does not
perform complex logical reasoning, it mostly assembles and integrates
evidence assemblies and instantiated theories.  This means that the
argument interprets claims as atomic propositions (i.e., it ignores
any internal structure).
The interpretation of each reasoning step in an argument is that the
conjunction of its subclaims implies its parent claim.  However, as
discussed earlier and
illustrated (in yellow) in Figure \ref{blocks-hand}, Assurance 2.0 blocks
generally have a \emph{side-claim} whose purpose is to ensure that the
subclaims are well formed and really do deductively entail the parent
claim.  The relationship is then 
\vspace*{-1ex}
$$\mbox{side-claim} \supset
(\mbox{conjunction of subclaims} \supset \mbox{parent
claim}),\ifelse{\footnote{We use $\supset$ for material implication, $\wedge$
for conjunction, $\neg$ for negation, $\equiv$ for equivalence, and
$\approx$ for approximate (numerical) equality.}}{}
\vspace*{-1.5ex}
$$
which is logically equivalent to
\vspace*{-1.5ex}
\begin{equation}\label{decomp-logic}
(\mbox{side-claim} \wedge \mbox{conjunction of subclaims}) \supset \mbox{parent claim},
\vspace*{-1ex}
\end{equation}
and so we see that although
the side-claim has a conceptually distinct status, it is logically no
different from the other subclaims.


Claims in Assurance 2.0 are written in natural language and \clasce\
does not attempt to interpret them\footnote{We have participated in
experiments where Large Language Models such as ChatGPT
translate natural language claims into a notation of
``\texttt{Object, Property, Environment}'' triplets that can be
subject to \emph{semantic analysis} using tools for Answer Set
Programming \cite{Murugesan-etal24:TDLP}, and we plan to develop
and apply this methodology.}  and cannot assess soundness
of reasoning steps.  Instead, this assessment requires human judgment
to interpret the claims and the narrative justifications supplied with
each step.  Rather than annotate each step to indicate a positive
assessment, \clasce\ assumes all steps are assessed sound and its
human users can indicate dissent by attaching a defeater node (see
Section \ref{defeaters}) to any step whose soundness is doubted.  This
allows soundness to be incorporated in the automated validity checks
for arguments with defeaters (again, see Section \ref{defeaters}).

A complete assurance argument is considered sound when it is logically
valid, human assessment concurs that the narrative justification
supplied with each reasoning step supports indefeasible confidence
that its parent claim is deductively entailed by its subclaims and
side-claim and, likewise, the justification for each evidence
incorporation step supports indefeasible confidence, corroborated by strong positive
confirmation measures, in its ``something useful'' claim.

Logical soundness is the most important assessment applied in
Assurance 2.0 as it means the argument supports indefeasible
confidence in the top claim.   However it lacks graduation, and for
that we turn to probabilistic assessment.

\vspace*{-1.5ex}
\section{Probabilistic Assessment}
\label{prob}

Suppose we have a sound case, then reduce its threshold for weight of
evidence and reduce the quantity or quality of evidence accordingly
(e.g., instead of testing to MC/DC coverage, we do only branch coverage);
the case remains sound, but we are surely less confident in its top
claim.  A different ``weakening'' is seen in DO-178C \cite{DO178C},
where Design Assurance Levels (DALs) A to C require both High and Low
Level Requirements (HLR and LLR), whereas (the lower) Level D requires
only HLR\@.  Intuitively, the idea is that we are less confident of
the large ``leap'' in reasoning from implementation directly to HLR
than of the combination of steps from implementation to LLR and then
to HLR\@.  This would be manifest in the Level D case as difficulty
justifying side-claims on the substitution between properties of the
implementation and those of the HLR.

The motivation for these ``weakened'' cases is that they should be
cheaper to produce, yet might still be adequate for less critical
systems, or for less critical claims.  Dually, we would like some
basis for believing that the additional cost of the original
``strong'' cases does deliver greater confidence in their claims.
What we seek, therefore, is a way to augment logical soundness with a
graduated measure that indicates the strength of our confidence in the
case.

Confidence is naturally expressed as a probability and there is a
valuable relationship between confidence in an assurance case
and inferences that can be made about probability of failure for the
system concerned \cite{Strigini&Povyakalo13}.  However, probability of
failure can also be made an internal part of the assurance argument:
that is, it can be stated in claims (e.g., ``system has $\texttt{pfd}
< 10^{-6}$\,'') and justified by evidence and reasoning 
(e.g., by reference to suitable theories for reliability estimation).
This is the best approach when explicit quantification is required.

On the other hand, when we merely wish to compare different arguments,
evidence, or theories, we could assess confidence externally as a
subjective holistic evaluation of the entire case\textbf{---}however, a more
principled method for evaluation of this kind is to calculate it as the
composition of assessments for the basic elements of the case,
including its evidence and individual reasoning steps.  This will
involve some combination of logic and probability, which is a
notoriously difficult topic as the semantics of the two fields
have different foundations \cite{Adams98}.

Nonetheless, there are numerous proposals for calculating
probabilistic confidence in assurance cases by methods of this kind;
however, a study by Graydon and Holloway cast doubt on many of them
\cite{Graydon&Holloway:quant17}.  
\longver{
Graydon and Holloway examined 12
proposals that use probabilistic methods to quantify confidence in
assurance case arguments: five based on Bayesian Belief Networks, five
based on Dempster-Shafer or other forms of evidential reasoning such
as J{\o}sang's opinion triangle, and two using
other methods.  By perturbing the original authors' own examples, they
showed all the proposed methods can deliver implausible results.

}
We suspect that the reason for this disappointing behavior is that the
methods concerned are attempting a double duty: they aim to evaluate
confidence in the case, but must also (implicitly) assure its
soundness.  Probabilistic methods are poorly suited to the latter
task, which is more naturally cast in terms of logic.  In Assurance
2.0 we separate these  and assess soundness as a logical
property, as described in the previous section, and only for cases
assessed as sound do we proceed to assess probabilistic confidence.
Nevertheless, we do intend to explore Graydon and Holloway's examples
when our tools are fully developed.

Our method for probabilistic assessment is compositional over the five
building blocks of Assurance 2.0 arguments and works bottom up: from
the evidence and assumptions at the leaves up to the top claim.  As we
explained in Section \ref{confirmation}, the subjective posterior
probability $P(C \vbar E)$ is a natural expression of confidence in
the claim $C,$ given the evidence $E.$ However, when assessing
soundness we use a confirmation measure rather than the posterior
probability because we wish to evaluate the discriminating power, or
``weight,'' of the evidence, and confirmation measures do this.  But
once we have assessed soundness, it is reasonable to use the posterior
as our measure of probabilistic confidence in the claim $C$ and it is
this that is propagated upward in probabilistic assessment of the
rest of the argument.  Note that $C$ should be the ``something
useful'' claim, not the measured one.  And note that in Section
\ref{confirmation} we condone informal (i.e., non-numerical)
interpretation of confirmation measures, but for confidence we do
require an actual numerical estimate for the subjective probability
$P(C \vbar E)$\@.  Developers must likewise assign a numerical
probability to assumptions, and should provide justification for their
choice (e.g., historical experience).

How probabilities should be assigned to residual doubts is a
delicate choice: if they are truly residual (see Section
\ref{residuals}), they can be ignored (as they are in logical
assessment), but developers may alternatively provide numerical
estimates for the purposes of exploration and analysis.

Once we have assessments for probabilistic confidence in the leaf
nodes, we can propagate numerical assessments upward.
For simplicity, we start by considering argument blocks that have only
a single subclaim, such as the generic substitution or concretion block
shown in Figure \ref{substconc}.  For logical soundness, the parent
claim $C$ must be deductively entailed by the subclaim $S$, subject to the
side-claim $W$.  Recall from (\ref{decomp-logic}) that this is
interpreted as $W \wedge S \supset C.$ We must now apply a
probabilistic interpretation to this implication, so that
\begin{eqnarray*}
\conf(C) & \approx & \conf(W \wedge S)\\
 & \approx & \conf(W) \times \conf(S \vbar W)
\end{eqnarray*}
where $\conf(x)$ denotes probabilistic confidence in claim $x$.

\begin{wrapfigure}{r}{.5\textwidth}
\ifelse{\vspace*{-6ex}}{\vspace*{-6ex}}
\begin{center}
\includegraphics[width=0.5\textwidth]{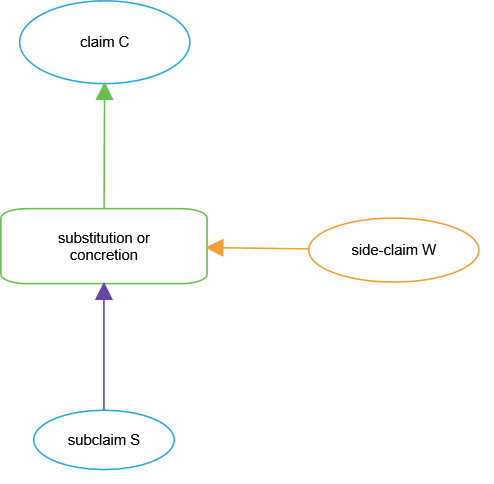}
\end{center}
\vspace*{-4ex}
\caption{\label{substconc}Substitution/Concretion Block}
\ifelse{\vspace*{-4ex}}{\vspace*{-4ex}}
\end{wrapfigure}

We expect the lower steps of the argument (i.e., the subcases
supporting $W$ and $S$) to supply $\conf(W)$ and $\conf(S)$\@, but
$\conf(S)$ is not the same as $\conf(S \vbar W)$ (unless $S$ and $W$
are independent), so this is not quite what is required.  However, the
structure of a sound assurance case is such that all the claims and
subclaims appearing in its argument must be true, so when we
evaluated the subclaims and evidence contributing to $S,$ we
implicitly did so in a context where $W$ is true.  Hence, our
assessment of probabilistic confidence in the subclaim $S$ is really
confidence \emph{given} the rest of the argument, and so the confidence we
labeled $\conf(S)$ is ``really'' $\conf(S \vbar W)$ and probabilistic
confidence in $\conf(W \wedge S)$ can indeed be taken as the product
of probabilistic confidence in its two subclaims.  Thus \[\conf^P(C) =
\conf^P(W) \times \conf^P(S)\] where the superscript $P$ in
$\conf^P(x)$ indicates this is the ``product'' calculation.

\longver{
\begin{figure}[h]
\center
\includegraphics[width=1.0\linewidth,trim=2.9cm 9cm 3cm 8.3cm, clip=true]{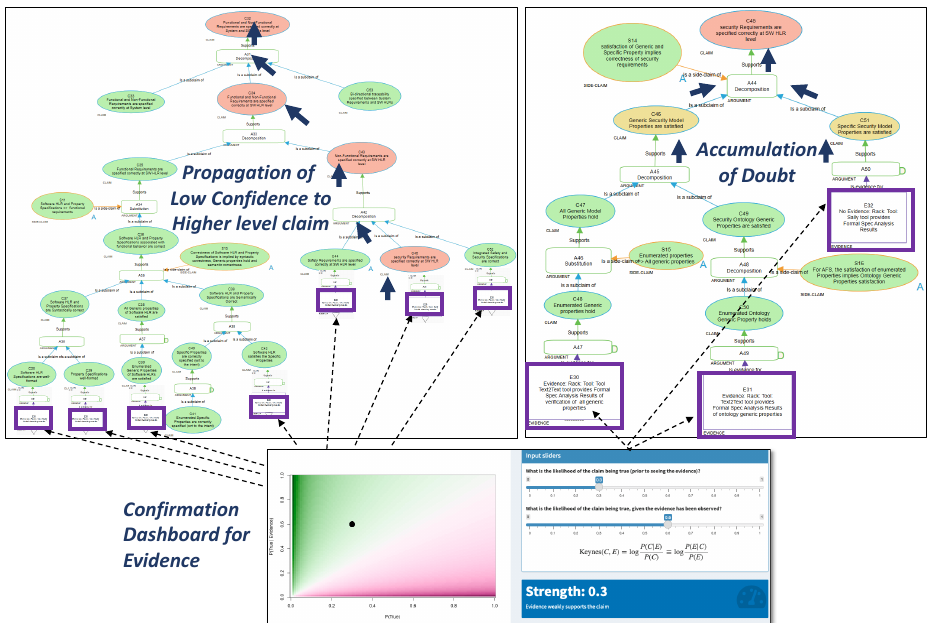}
\vspace*{-4ex}
\caption{Illustration of Confirmation Measures and Confidence Propagation (from \cite{Clarissa24:Sassur})}
  \label{widgets}
\vspace*{-4ex}
\end{figure}
}

Some may feel the ``really'' assumption in this calculation is too
aggressive and would prefer a more conservative approach.  One such is
``sum of doubts''\,: our doubt in the parent claim is no
worse than the sum of doubts for its subclaims and side-claim
\cite{Adams98}.  The ``doubts'' referred to here are
\emph{probabilistic doubts} as opposed to the use of the term in the 
section that follows, where it
means general disquiet or concern.  Probabilistic doubt in a claim $x$
is probabilistic confidence in its negation $\conf(\neg x),$ which is
$1 - \conf(x)$, and so 
\vspace*{-0.7ex}
\begin{equation}\nonumber
\conf^D(C) \geq \conf^D(W) + \conf^D(S) - 1
\end{equation}
\vspace*{-3.5ex}

\noindent where the superscript $D$ indicates this is the ``sum of
doubts'' calculation.  Generalizing these derivations
for decomposition and calculation blocks, which may have $n$ subclaims $S_1,\ldots,S_n$  as shown in Figure
\ref{blocks-hand}, we
obtain
\begin{eqnarray*}
\conf^P(C) & = & \conf^P(W) \times \prod_{i=1}^n \conf^P(S_i)\ \mathrm{and}\\
\conf^D(C) & = & \conf^D(W) + \sum_{i=1}^n \conf^D(S_i) - n.
\end{eqnarray*}

\ifelse{ 

\clasce\ is able to perform these propagations and its users can also
make manual adjustments (see \cite{Bloomfield&Rushby:confidence22} for
examples and additional details).  Note, however, that by structural
induction, we can see that the confidence propagated by the product
calculation to any interior node is simply the product of the
confidence at all leaves of its subtree, and likewise its
probabilistic doubt is simply the sum of the doubts at those
leaves.  This means that if confidence remains the same at the leaves,
then changes to the ``shape'' and content of the argument will have no
effect on probabilistic confidence at the top node.  Furthermore,
excising any subtree (i.e., removing evidence and some part of the
argument) will \emph{increase} probabilistic confidence at the top
node.

But the point is that confidence at the leaves will not remain the
same under these manipulations: confidence in evidential nodes will
change as evidence changes or is required to justify different claims,
and an excised subtree will surely cause the reasoning step to which
it was attached to become nondeductive.  Thus, changes to an argument
will compel further adjustments to restore deductiveness and
soundness.  Propagation of probabilistic confidence, particularly
where human judgment overrides the automated calculation, provides a
tool for understanding the impact of changes to an argument and
thereby allows the relative strengths of different arguments for the
same system to be explored and documented in an explicit and rational
way.  We stress, however, that the absolute values of the
probabilities propagated carry little significance: their purpose is
to support tradeoffs of \emph{relative} confidence versus effort and
cost, as needed when developing graduated forms of assurance for
different levels of risk \cite{Daw-etal:levels-DASC23}, and it also
supports deliberate allocation of effort across a single assurance
case.  } { \clasce\ is able to perform and provide visualizations of
these propagations (illustrated in Figure \ref{widgets})
and its users can also make manual adjustments (see
\cite{Bloomfield&Rushby:confidence22} for examples and additional
details).  We stress that the absolute values of the probabilities
propagated carry little significance; the purpose of these
calculations is to allow the relative strengths of different arguments
for the same system, and different parts of the same argument, to be
explored and documented in an explicit and rational way.  This
supports tradeoffs of effort and cost versus confidence, as needed
when developing graduated forms of assurance for different levels of
risk \cite{Daw-etal:levels-DASC23}, and it also supports deliberate
allocation of effort across a single assurance case.

Assessments of tradeoffs for graduated assurance may not
be simple, however.  By structural induction, it is easy to see that the
confidence propagated by the product calculation to any interior node
is simply the product of the confidence at all the leaves of its
subtree, and likewise its probabilistic doubt is simply the sum of the
doubts at all those leaves.  This means that if confidence remains the
same at the leaves, then changes to the ``shape'' and content of the
argument will have no effect on probabilistic confidence at the top
node.  Furthermore, excising any subtree (i.e., removing evidence and
some part of the argument) will \emph{increase} probabilistic
confidence at the top node.  These counterintuitive observations might seem to reduce
probabilistic confidence to irrelevance or falsehood, but this is not
so: for confidence at some of the leaf (particularly evidential) nodes
will surely change as they are required to justify different claims,
and human judgment may override the default propagations at some
interior nodes as the content of the reasoning steps is modified to
accommodate a changed ``shape'' to the argument.  (And note that if a
subtree is excised, the reasoning step to which its root was a
subclaim can no longer be deductive---unless that subtree was
redundant---so modifications will be required in the argument above the
excised subtree.)

Thus, changes to an argument that are intended to reduce its cost
(e.g., by eliminating or simplifying evidence) will compel further
adjustments to restore logical deductiveness and soundness, and human
judgment must assess confidence in the changed evidence and reasoning
steps.  Propagation of probabilistic confidence, particularly where
human judgment overrides the automated calculation, provides a
rational tool for understanding and accumulating---and experimenting
on---the impact that these changes have in confidence at the top
claim.  In future work, we hope to use this approach to examine the
example developed by Daw, Beecher and Holloway
\cite{Daw-etal:levels-DASC23}

}

\textbf{Note (August 2025):} after this paper was published, we
developed new methods of probabilistic assessment for Assurance 2.0
that do pay attention to the ``shape'' of the argument and, in
particular, to the reasons why subclaims of a decomposition block
support its parent claim \cite{Bloomfield&Rushby:Probconf25}.  The new
methods are significantly more precise than the ``product'' and ``sum
of doubts'' methods described here and consequently we no longer
recommend these methods for general use.

\section{Defeaters and Dialectical Exploration}
\label{defeaters}

\memo{Somewhere need to acknowledge doubts could flow down as well as up.}

We have described indefeasible logical soundness as our primary method
for assessing confidence in an assurance argument, with probabilistic
assessment as a secondary method that provides a way to graduate the
degree of confidence.  In addition to these ``positive'' perspectives,
we also need to be sure there are no overlooked or unresolved doubts
that could change the judgment and we refer to investigation of
doubts as the ``negative'' perspective on
assessment.

We refer to any concern about an assurance case as a \emph{doubt} and
we record it by adding a \emph{doubt node} to the graphical
representation of the assurance argument, pointing to a node that is
under suspicion.  The doubt node contains a claim indicating the
nature of the doubt (e.g., ``I think there is something wrong here'').
At some point, we must return to investigate the nature and origin of
the doubt and will either dismiss it as unwarranted, or refine and
sharpen it into a \emph{defeater} with a more specific claim (e.g.,
``the justification for this step is inadequate'') whose investigation
and outcome (i.e., confirmed or not) are recorded in a subcase
attached to the defeater.  Thus, a doubt is simply a defeater that has
not yet been investigated (i.e., has no subcase) and so we generally
refer to both as defeaters.  The back and forth investigation of an
assurance case argument against doubts and defeaters is an application
of the \emph{Socratic} or \emph{dialectical} methods for exposing
error and refining beliefs.  These date back to
ancient Greece but retain their potency.  In particular, defeaters
play a r{\^o}le in argument that is similar to falsification in
science and mathematics \cite{Lakatos}; they can also be seen as the
analog, for arguments, of hazards to a system.  Thus, identification
of potential defeaters should not be seen as criticism but as a
contribution to the development and clear formulation of an assurance
case and part of a process to establish its indefeasibility.  In
addition, developers should consciously generate doubts and should
vigorously investigate their associated defeaters as a guard against
confirmation bias, and evaluators may raise potential defeaters as a
way to elicit additional explanation or to clarify their understanding
of some part of an assurance case.

The mere act of pointing a defeater at a node means that the argument
can no longer be assessed as logically valid or sound, but that
judgment may be refined when the defeater is supplied with a
subargument that shows its claim to be definitely \texttt{true} or
\texttt{false}.  
If the defeater is supported by an assurance subargument that is
adjudged to be sound, so that its claim is \texttt{true}, then the
defeater is said to be confirmed or \emph{sustained} and the main
argument, and possibly the system it is about, must be modified to
overcome the flaw that has been identified.  After these
modifications, the defeater and its subargument should no longer
apply, but we might like to retain them in the case as documentation
to assist future developers and evaluators.  Thus, a defeater can be
marked \emph{addressed} and it and its subcase will then be treated as
a comment.  Because the defeater no longer applies to the now modified
primary argument, a narrative description of the original problem and
its resolution should be added to the defeater node.  But this may be
difficult to understand because the context has changed from the
original to the
modified argument, so another choice is to alter the previously
sustaining subcase for the defeater into a refuted subcase (see below)
against the modified primary argument.  An alternative response to a
sustained defeater, provided the identified flaw is judged 
insignificant, is to explicitly accept it as a \emph{residual doubt}
(see Section \ref{residuals}).

If we suspect that a defeater is a ``false alarm,'' or it is one that
has been overcome by modifications to the original case (as above),
then our task is to \emph{refute} it: that is, to provide it with a
subcase that shows it to be \texttt{false}.  One way to do this is
with a second-level defeater that targets the first defeater or some
part of its subargument.  (Another way is by use of
\emph{counter-evidence}, see \cite{Bloomfield-etal:defeaters24}.)  If
the assurance subcase for that second-level defeater is sustained,
then the first defeater is said to be \emph{refuted} and it and its
subcase play no part in the interpretation of the primary case, but
can be retained as commentary to assist future developers and
evaluators who may entertain doubts similar to that which motivated
the original defeater.

The introduction of refutational arguments means that our notion of
logical validity needs to be enriched.  Previously (in Section
\ref{logic}), we propagated validity upward from the leaves of the
argument and, implicitly, we categorized its claims as either
\texttt{true} or \texttt{unsupported}.  Now we need to add a third
category, \texttt{false}, and must develop rules for propagating these
three values.  It might seem that we could look to the methods of
defeasible reasoning or nonmonotonic logic for this purpose, but the
goal of these methods is to work out what can be concluded when there
are contradictory premises or when exceptions are added, whereas, in
Assurance 2.0, our goal is to determine which parts of a case
\emph{are} contested: that is, called into question by defeaters,
possibly at several levels.

We do not interpret the claims in an assurance argument when assessing
validity, and the same applies to claims made by defeaters.  In
particular, if a defeater with claim $X$ pointing to a claim $A$ is
sustained, we do not suppose that some logical combination of $A$ and
$X$ is thereby justified; we accept that the claim $A$ is challenged
and revise it and/or its supporting subcase to overcome the source of
doubt.  Of course, we must make the human judgment that $X$ has some
impact on the credibility or relevance of $A$ but we do not reduce
this to some logical requirement such as $X \equiv \neg A$.  Having
said that, in Section \ref{elim} we will introduce a circumstance
where we do recognize the special case where the claim in a defeater
is the negation of that in the node that it points to; we call these
\emph{exact} defeaters (the general kind are then known as
\emph{exploratory} defeaters).

We now develop the propagation rules for validity in the presence of
defeaters.  Since a defeater can point to any kind of node, we define
the claim \emph{affected by} the defeater to be the node pointed to if
this is a claim, assumption, residual doubt, or defeater, and
otherwise the parent claim (which may be a defeater) of the node
pointed to.

We first consider propagation from a defeater to  its affected claim.
When the claim in an exploratory defeater is assessed \texttt{false}
it means the defeater is refuted; hence, the main argument (or subargument for
lower-level defeaters) is exonerated and its claims are assessed as if
the defeater were absent.  When the claim in a defeater is assessed
\texttt{unsupported} (which also applies when the defeater has no
subcase\textbf{---}i.e., it is merely a doubt), then so is the claim affected
by the defeater.  And when the claim in the defeater is assessed
\texttt{true}, then the affected claim is also assessed
\texttt{unsupported}; it cannot be assessed \texttt{false} because the
defeater may not precisely refute the affected claim (unless it is an
exact defeater, which is considered later), but merely call it into
question.

These assessments override those due to any other nodes pointing to
the affected claim (which may affirm it as \texttt{true}): when a
claim is challenged by a \texttt{true} or \texttt{unsupported}
defeater, we have to accept that it is called into question.  However,
the appropriate response may require further diagnosis.  When the claim
affected by a defeater is assessed as \texttt{unsupported}, we need to
examine the assessment of its defeater (\texttt{unsupported} or
\texttt{true}) to determine the response: in the former case, the
defeater's subcase needs more work, while in the latter the main
argument needs to be revised (and possibly also the system concerned).

Finally, we consider propagation of assessments through reasoning
steps; recall, in NLD, individual reasoning steps are intended to be
deductively valid and are interpreted as material implications of the
form shown in formula (\ref{decomp-logic}).
The sideclaim and subclaims constitute the \emph{antecedent} to this
implication.  As described in Section \ref{logic}, when all claims in
the antecedent are assessed \texttt{true} then, by the rules of
classical logic, so is the parent claim.  And if any antecedent claims
are \texttt{unsupported}, then the parent claim is also.  But suppose
some claims in the antecedent are assessed \texttt{false}.  Since they
are conjoined,  the whole antecedent becomes \texttt{false}; does
this mean we should assess the parent claim as \texttt{false} too?

It does not: it would be attempting to derive $\neg A \supset \neg B$
from $A \supset B$, and this is the logical fallacy of ``denying the
antecedent.''   Moreover, there is a
further problem: if the antecedent is \texttt{false}, then it can
imply any parent claim (this is the ``false implies everything''
problem).  Thus, in general, we cannot propagate \texttt{false} upward
through reasoning steps; we must do something weaker and the appropriate response is
to assess the parent claim as \texttt{unsupported}.

It is long standing practice to challenge and review assurance cases,
but the explicit use of defeaters to systematize and record these
activities is new, and the pragmatics how best to postulate, examine,
respond to, and manage them needs further evaluation and tool support.
\clasce\ has a ``validity plugin'' that can evaluate validity of an
argument in the presence of defeaters using the method described
above.  It allows developers and evaluators to see which defeaters
have been refuted and which are still active, and which parts of an
argument are called into question by the active defeaters.  This is
intended to support dialectical exploration of assurance arguments, so
that confidence can be probed and ultimately more firmly established.
In addition, we are currently exploring conceptual sources
of defeaters and systematic, and potentially automated, ways to
generate useful classes of defeaters.

Defeaters introduce refutational reasoning to Assurance 2.0 and this
allows an alternative form of assurance argumentation, as we now
describe.

\vspace*{-1ex}
\subsection{Exact Defeaters and Eliminative Argumentation}

\label{exact}

\label{elim}

``Eliminative Induction'' is a method of reasoning that dates back to
Francis Bacon who, in 1620, proposed it as a way
to establish a scientific theory by refuting all the reasons why it
might be false (i.e., all its defeaters).Weinstock, Goodenough, and Klein \cite{Goodenough-etal:ICSE2013} build
on the idea of Eliminative Induction to develop a means of assurance
that they call \emph{Eliminative Argumentation}.  Here, instead of
attempting to confirm a positive claim such as ``the system is safe''
we instead attempt to refute the negative claim ``the system is
\emph{un}safe.'' A successful refutation will establish the negation
of that claim, namely ``the system is \emph{not} unsafe.''  In
classical (as opposed to intuitionistic) logic this establishes the
positive claim by virtue of the rule for elimination of double
negation, and thereby provides the desired assurance.  Millet and
colleagues report successful application of eliminative argumentation
in assurance of real systems \cite{Millet:CERN-LHC23}.

The basic methodology of Assurance 2.0 supports development of
\emph{positive cases} where a constructive argument is developed in
support of some beneficial claim about a system.  Nonetheless, we
explicitly introduce defeaters and confirmation measures to 
help address complacency and bias by inviting consideration of
contrary points of view.
Furthermore, we recognize that it can sometimes be useful to consider
fully contrary or counter-claims, counter-evidence and counter-arguments, and we
introduce \emph{exact} defeaters for this purpose; effectively, they
allow us to introduce negation into an assurance argument.

An exact defeater is one that: a) points to a node that is either a
claim or another defeater that b) lacks a subcase, and c) whose own
claim is the negation of the one pointed to.  Because claims in
\clasce\ are written in natural language, it is not trivial to
determine if one claim is the negation of another.  Accordingly,
\clasce\ provides an explicit selection in its interface to indicate
that a defeater should be treated as the exact negation of the claim
or defeater that it points to.  Furthermore, the node pointed to may
have a subcase, but it will be ignored (and indicated so in the
graphical presentation) when the node becomes the target of an exact
defeater.  This is to support exploratory development of a case
without having to undo or redo previous work.

The propagation rules for exact defeaters are simple: if the exact
defeater is assessed \texttt{unsupported}, then so is the node that it
points to; otherwise the assessment of the node pointed to is the
logical negation of the assessment of the claim in the defeater.

In the framework of Assurance 2.0, exact defeaters allow us to construct
\emph{negative (sub)cases}, and eliminative argumentation can then be
represented by attaching an exact defeater to a positive claim and
attempting to refute it.  Notice that whereas exploratory defeaters
\emph{augment} the main argument by providing an exploratory
investigation or commentary, exact defeaters are used as a reasoning
step \emph{within} the main argument.  Owing to the refutational
context in negative cases, it can be useful to have \emph{disjunctive}
decomposition blocks and \clasce\ supports these (see
\cite{Bloomfield-etal:defeaters24} for details).

\vspace*{-2ex}
\section{Assessment of Residual Doubts and Risks}

\label{residuals}

An assurance argument may contain residual doubts: these are
explicitly marked defeaters that we have been unable or have chosen
not to eliminate or fully mitigate.  In assessing logical soundness
and probabilistic confidence in an assurance case, we assume the
consequences of residual doubts are insignificant and, on that basis,
we ignore them.  We thereby incur an obligation to justify this
assumption.  In particular, we must consider the potential impact that
a faulty assessment could have on failure (i.e., defeasibility) of
the case.  Specifically, for each residual doubt, we must show that
the likelihood of wrongly assessing it (as residual), combined with
its worst possible consequences (i.e., its \emph{risk}), is
below some threshold for concern.

Residual doubts may be due to uncertainty in the environment: for
example, the system may be designed to withstand a single sensor
failure and historical evidence indicates this is sufficient, but it
is always possible to encounter more.  Or they may be due to
limitations of human review (e.g., human requirements tracing cannot
be guaranteed to be free of error), or to limitations in automated
analysis (e.g., automated static analysis may be unable to discharge
some proof obligations, leading to possible false alarms that must be
reviewed by humans, a potentially error-prone process).

\ifelse{\newpage}{}
In Assurance 2.0, we categorize residual risks into four levels of
severity.\ifelse{\\[-3.0ex]}{}
\begin{description}\itemsep=0.5ex

\item[Significant:] an individual residual doubt poses a risk that is
    above the threshold for concern.  In this case, the issue cannot
    be considered a merely ``residual'' risk, but must treated as a
    defeater and either eliminated or mitigated.

\item[Minor:] an individual residual doubt poses a risk that is below
    the threshold for concern, but it is possible that many such might
    cumulatively exceed the threshold.  An example could be static
    analysis, where we use human review to evaluate proof obligations
    that the automation cannot decide.  These risks need to be managed
    explicitly: 10 might be OK, but not 100.

\item[Manageable:] a class of minor residual risks whose number and
cumulative severity are below the threshold of concern.

\item[Negligible:] multiple residual doubts of a similar kind
    collectively pose a risk that is below the threshold for concern.
    This may arise when the source of doubt occurs many times but
    is adjudged to be trivial.  An example (depending on
    local policy) might be ``style'' warnings from a static analyzer.

\end{description}

\noindent At final assessment, the only residual doubts remaining should be
those whose risks are categorized negligible and those categorized
minor but manageable.

\vspace*{-1ex}
\section{Summary and Conclusion}

We have described methods for gaining and assessing confidence in
assurance cases based on Assurance 2.0 and its automated assistance
with \clasce.  
Here, we summarize these methods and provide brief
conclusions.  We do not provide references: they can be found in
earlier sections specific to each topic.

There is no simple approach that avoids the need for the essentials of
an assurance case: we cannot just say ``it has been proved correct,''
or ``it has been tested.''  We need confidence in correctness
\emph{together} with testing or operational experience, and we need to
articulate the theories behind their respective claims, how they have
been applied, how they fit together, and whether their assumptions are
and will continue to be valid.  Assurance 2.0 builds on earlier
treatments of assurance cases and their arguments, but is more rigorous
and demanding.  We claim that this simplifies their development and
assessment because issues that were previously treated in an \emph{ad
hoc} manner and subject to contention and challenge are now made
explicit and treated systematically.

In particular, we are explicit that the goal of an Assurance 2.0
argument is indefeasible justification, meaning we must be confident
there are no overlooked or unresolved doubts that could change its
assessment.  As consequences of this, reasoning steps are selected
from just five building blocks and must be deductive (or explicitly
acknowledge and manage the doubt if not).  Similarly, evidence is
weighed very deliberately using confirmation measures and we
distinguish carefully between facts established by the evidence
(claims about ``something measured'') and inferences drawn from it
(claims about ``something useful'').

These rigorous requirements and other supporting constraints enable
our primary positive assessment for an Assurance 2.0 argument to be
logical soundness, whose evaluation is straightforward.  Note that we
say it is straightforward, meaning it is clear what must be done, not
that it is easy: it requires expert technical judgment, but this
judgment can focus on technical issues without being distracted by
unmanaged doubts and contested interpretations.  Specifically, human
assessment must concur that the narrative justification supplied with
each reasoning step supports indefeasible confidence that its conjoined
subclaims and side-claim deductively entail its parent claim.
Similarly, the justification supplied with evidence must provide
indefeasible confidence, corroborated by strong positive confirmation
measures, in its ``something useful'' claim.

Logical soundness is the most fundamental assessment for an Assurance 2.0
case: it tells us that the argument and its evidence truly do support
the top claim, but it does not tell us how strongly they do so.  We
recommend that if the top claim needs to be quantified
probabilistically, then this should be stated in the claim and developed
as an internal part of the argument.  However, it can also be useful
to perform an external assessment of the relative strength of arguments,
and of parts of arguments, and we define a simple method for
probabilistic assessment of this kind.  The method operates
compositionally using either a ``product'' calculation or, more
conservatively, a ``sum of doubts.''  The numerical valuations are of
limited absolute significance, but they serve to explore the risk
of residual doubts and the relative strengths of different arguments
for the same system.  This allows rational tradeoffs of effort and
cost versus confidence, which is needed in developing graduated forms
of assurance for different levels of risk, as exemplified by the DALs
of DO-178C.  It also allows conscious apportionment of effort across
the different parts of a single argument.

While building a forceful positive case, the developers of an
assurance case must guard against complacency and confirmation bias.
This can be assisted by vigorous and active exploration of challenges
to, and doubts about, the case.  In Assurance 2.0, doubts are refined
and recorded as defeaters, which are nodes in the graphical
representation of the argument that explicitly challenge other nodes
and that may have their own subcases to sustain or refute them.
Sustained defeaters require revision to the assurance case and
possibly the system itself.

In addition to guarding against confirmation bias, the record of
doubts explored as defeaters assists assessors of the case.  When
previously examined defeaters are recorded as part of the case,
assessors may find that their own questions and doubts have been
anticipated and answered, thereby streamlining their task and also
enabling a constructive dialectical examination of the case.

The presence of defeaters requires that we keep track of which
defeaters are sustained, which have been refuted (e.g., by themselves
being defeated), and which are still under investigation.  We also
need to keep track of which parts of the overall argument are under
challenge by active defeaters.  We do this by extending the notion of
logical validity to arguments with defeaters and this also supports an
alternative approach to assurance by ``eliminative argumentation.''

All identified defeaters should be examined and resolved.  However, a
conscious decision may be made to accept some as residual doubts.  For
example, a subcase that uses testing to justify absence of runtime
exceptions may have residual doubt due to incompleteness of testing.
The risks posed by such doubts must be assessed and only those
categorized as manageable, and those categorized as negligible
can be allowed to remain as residual risks: others must be eliminated
or mitigated by revisions to the argument or the system.  The
probabilistic valuation of \clasce\ can be used to help visualize the
potential impact of residual doubts on the overall argument.

In conclusion, Assurance 2.0 assesses confidence in an assurance case
by considering both positive and negative perspectives.  The positive
perspectives are logical soundness and (optionally) a probabilistic
assessment; the negative perspectives are dialectical exploration of
potential defeaters, and careful evaluation of all residual doubts.
During development and, optionally, during assessment both positive
and negative aspects may be explored simultaneously, but at the
conclusion of both development and assessment, all potential defeaters
should have been dismissed, or accepted as residual risks, and the
positive perspective should deliver the judgment of indefeasible logical soundness.

We advocate that assurance cases should largely be synthesized from
instantiations of generic subcases for standard assurance topics that
we call \emph{assurance theories}.  Our hope is that future generations of
assurance guidelines and standards can be supplemented or replaced by
community-driven development of Assurance 2.0 theories, pre-assessed
by the methods described here.

\vspace*{-3ex}
\subsubsection*{Acknowledgments.}

The work described here was developed in partnership with other
members of the {\sc Clarissa} project, notably Srivatsan Varadarajan,
Anitha Murugesan and Isaac Hong Wong of Honeywell, Gopal Gupta of UT
Dallas, and Kate Netkachova and Robert Stroud of Adelard.

This material is based upon work supported by the Air Force Research
Laboratory (AFRL) and DARPA under Contract No. FA8750-20-C-0512.  Any
opinions, findings and conclusions or recommendations expressed in
this material are those of the author(s) and do not necessarily
reflect the views of the Air Force Research Laboratory (AFRL) and
DARPA.

\ifelse{\vspace*{-3ex}}{}

\bibliographystyle{splncs04}

\end{document}